\documentclass[twocolumn,openany,openright,amsmath,amssymb,superscriptaddress,prb]{revtex4-1}

\usepackage{float}
\usepackage{latexsym} 
\usepackage{amsmath,amsthm}
\usepackage{ifpdf}
\usepackage{epstopdf}
\usepackage{subfig}
\usepackage{soul}
\usepackage{dcolumn}
\usepackage{bm}
\usepackage{braket}
\usepackage{wrapfig}

\usepackage{color}

\usepackage{graphicx}

\usepackage[final]{changes}

\newcommand {\Pdert}{\frac {\partial}{\partial t}}

\newcommand{\haq}{\hat a_\qq}
\newcommand{\adq}{\hat a^\dagger_{\qq}}

\newcommand{\hbq}{\hat b_{-\qq}}

\newcommand{\bdq}{\hat b^\dagger_{-\qq}}

\newcommand{\kk}{\mathbf{k}}

\newcommand{\qq}{\mathbf{q}}

\newcommand{\vv}{\mathbf{v}}
\newcommand{\rr}{\mathbf{r}}

\newcommand{\vett}[1]{\mathbf{#1}}
\newcommand{\uvett}[1]{\hat{\vett{#1}}}

\begin{document}
\def \k{\bold k}
\def \l{\bold l}
\def \i{\bold i}
\def \p{\bold p}
\def \r{\bold r}
\def \qq{\bold q}
\def \A{\bold A}
\def \beq{\begin{equation}}
\def \eeq{\end{equation}}
\def \beal{\begin{aligned}}
\def \eal{\end{aligned}}
\def \bes{\begin{split}}
\def \ees{\end{split}}
\def \besu{\begin{subequations}}
\def \esu{\end{subequations}}
\def \g{\gamma}
\def \G{\Gamma}
\def \ac{\alpha_c}
\def \barr{\begin{eqnarray}}
\def \earr{\end{eqnarray}}

\title{Nonlinear optical response of Type-II Weyl fermions in two dimensions}
\author{Yaraslau Tamashevich}
\affiliation{Faculty of Engineering and Natural Sciences, Tampere University, Tampere, Finland}
\author{Leone Di Mauro Villari}
\affiliation{Department of Physics and Astronomy, University of Manchester, Manchester M13 9PL, UK}
\author{Marco Ornigotti}
\email{marco.ornigotti@tuni.fi}
\affiliation{Faculty of Engineering and Natural Sciences, Tampere University, Tampere, Finland}

\begin{abstract}
We present a theoretical model to study the nonlinear optical response of materials with a low-energy spectrum characterised by strongly tilted, type-II, Weyl cones in two dimensions, which we call 2D Weyl materials. Our findings reveal that the tilted nature of the Weyl cones is responsible for the appearance of even harmonics in the nonlinear signal, as well as its strong polarisation dependence. We discuss how it is possible to control such nonlinear response and envision how 2D Weyl materials can be used to realise novel photonic devices for sensing applications.
\end{abstract}

\maketitle

\section{Introduction}
Relativistic massless and massive fermions can be probed with high-energy physics experiments, but also appear as low-energy quasi-particle excitations in condensed matter systems, where their massless character is typically protected by crystal symmetries. The low energy dispersion of such excitations usually contain diabolical points and conical intersections, that are frequently referred to as Dirac, or Weyl, points. A prime example of that is certainly graphene. Since its discovery in 2004 by Novoselov and Geim \cite{ng,mam}, in fact, condensed matter physics has been witnessing a rapid expansion in the fabrication and engineering of a wide variety of materials with pseudo-relativistic quasiparticle excitations: the so-called Dirac materials \cite{wb}. This class of materials encompasses, amongst others, semimetals \cite{thakur_dynamic_2018}, transition-metal-dichalcogenides (TMDs) \cite{xu_graphene-like_2013}, and topological insulators \cite{hasan_colloquium_2010}. 
A considerable amount of effort has then been made in the last decade to understand the electronic and optical properties of such materials, and their underlying fundamental physical principles, both at the linear \cite{Jia_2019} and nonlinear \cite{Guo_Xiao_Wang_Zhang_2019,You_Bongu_Bao_Panoiu_2018} level. The research in this field, for example, has paved the way for breakthrough in spintronics and valleytronics \cite{sh}, generation of strong harmonics \cite{Kumar_Kumar_Gerstenkorn_Wang_Chiu_Smirl_Zhao_2013, Wang_Chien_Kumar_Kumar_Chiu_Zhao_2014, Ornigotti_Ornigotti_Biancalana_2021}, unconventional exciton states \cite{bgb,bb}, superconductivity \cite{cf}, but also the discovery of new topological states of matter \cite{hasan_colloquium_2010}.The universality of these concepts, moreover, very rapidly contaminated many others fields of physics, leading very quickly to, e.g., the fields of topological mechanics \cite{Beenakker_Kouwenhoven_2016}, topological photonics \cite{Lu_topological_2014, Ozawa_2019}, and topological atomic physics \cite{Goldman_Budich_Zoller_2016, Dalibard_Gerbier_2011}.

To understand what are the properties and peculiarities of the various Dirac materials, one could first try to classify them based on the nature of their low-energy dispersion, namely if they admit Dirac or Weyl (DW) cones. 
%
%
Dirac materials contain both time reversal and spatial inversion symmetry. When one of these symmetries is broken, the Dirac points are split into two constituent Weyl points, and the  medium becomes a Weyl material. In bulk materials both Dirac and Weyl points have been observed \cite{amv}. In two dimensions,
%
the amount of research on Dirac and Weyl fermions has been remarkable \cite{wb,ng1}. However, in contrast to Dirac fermions, which are easy to observe and engineer in 2D materials, two dimensional Weyl fermions have been only explored theoretically and, although several materials have been suggested, they are still eluding experimental verification \cite{gz,cs,Lin}. 

DW cones (fermions) can be further classified into 
Type I and Type II cones. While the former present DW cones that are straight, and preserve Lorentz invariance, the latter are tilted and with broken Lorentz invariance.
%
%
%
Type II cones typically occur when Type I DW nodes are tilted enough, along some specific direction, that a Lifschitz transition occurs and the system acquires a finite density of states at the DW node. The first proposal of Type II Weyl fermions \cite{sg} has inspired intensive investigations of the counterpart Type-II Dirac fermions, to the point that during the last few years a good deal of research has been done to characterise and find materials showing tilted DW cones. Type II fermions have been found in a variety of materials such as semimetals, TMDs (PtTe$_2$, WTe$_2$) \cite{my}, LaAlO$_3$/ LaNiO$_3$/ LaAlO$_3$ quantum wells \cite{tt} and PdTe$_2$ superconductors \cite{nj}.

 While the electronic properties of such media are fairly well studied, understanding the nonlinear optical processes, and harnessing their potential applications for photonics, constitutes a less explored territory. This work envisions the characterisation of the nonlinear optical properties of  materials with a low energy spectrum described by strongly tilted (Type-II)  Weyl cones in two dimensions, which we call 2D Weyl materials (2D-WMs). In particular we are interested in systems with band-crossing nodal points (massless fermions) and broken inversion symmetry such as Cr$_2$C and XP$_2$ transition metal diphosphides in which Type-II Weyl fermions are predicted \cite{meng,Aut}.  
 
 This work is organised as follows: in Sect. II we present the generalised model for 2D-WMs in the low-energy approximation. Section III is then dedicated to analysing the temporal and spatial symmetries possessed by such materials, and we show explicitly, how the inversion symmetry is broken. Section IV is then dedicated to obtaining Dirac-Bloch equations for 2D-WMs, which are then used in Sect. V to construct the nonlinear induced current. In Sect. VI, we discuss the properties of the harmonics generated by an electromagnetic pulse interacting with 2D-WMs, and, in particular, we point out the role the tilting od DW cones plays in said nonlinear response. Finally, conclusions are drawn in Sect. VII.

\section{Hamiltonian for a 2D Weyl Material}
To start our analysis, we consider the following low-energy, generalised, two-band Weyl Hamiltonian in $d$ dimensions
\beq \label{Hty2.1}
H_\xi(\qq) = \sum_{\mu=0}^3 c^\xi_{\mu,\qq} \sigma_\mu,
\eeq
where $\sigma_{\mu}$ are the Pauli matrices (with $\sigma_0=\mathbb{I}$), 
$\xi$ indicates a nodal point with opposite chirality (valley degree of freedom), and the coefficients $c_{\vett{q},\mu}^{\xi}$ are linked to the velocity and tilting parameters of the 2D-WM as follows:
\besu\label{cCoefficients}
\begin{align}
    c_{\vett{q},0}^{\xi}&=\vett{a}\cdot\vett{q}_{\xi},\\
    \left( c_{\vett{q},1}^{\xi}\, c_{\vett{q},2}^{\xi}\right)&=\left(\uvett{v}\cdot\vett{q}_{\xi}\right)^T,\\
    c_{\vett{q},3}^{\xi}&=\xi v\,q_x+\frac{\Delta}{2},
\end{align}
\esu
where $\vett{a}=(a_x,a_y)$ is the tilting vector, describing the orientation and magnitude of the tilted cone, $\vett{q}_{\xi}=(\xi q_x,q_y)$ is the valley-dependent momentum, $\hat{\vv}$ is carrier velocity tensor, whose diagonal elements $v_{ii}$ represent the carrier velocity along the direction $x_i$, and $v$ and $\Delta$ are, respectively, the inversion symmetry breaking parameter and a staggered potential accounting for an eventual gap in the band structure. If we introduce the following set of generalised polar coordinates
\besu\label{polarCoord}
\begin{align}
    g_{\vett{q}}^{\xi}&=\sqrt{\left(c_{\vett{q},1}^{\xi}\right)^2+\left(c_{\vett{q},2}^{\xi}\right)^2},\\
    \varphi_{\vett{q}}^{\xi}&=\arctan\left(\frac{c_{\vett{q},2}^{\xi}}{c_{\vett{q},1}^{\xi}}\right),
\end{align}
\esu 
the Hamiltonian in Eq. \eqref{Hty2.1} can be written in matrix form as follows 
\beq \label{heff1}
 H_\xi(\qq) = \left(\begin{array}{cc} c^\xi_{\qq,0}+c^\xi_{\qq,3} & g^\xi_\qq e^{-i\varphi^\xi_\qq}  \\ \ g^\xi_\qq e^{i\varphi^\xi_\qq} &c^\xi_{\qq,0}-c^\xi_{\qq,3} \end{array}\right).
\eeq
The form of the above Hamiltonian is particularly convenient and is readily diagonalised, with eigenstates and eigenvalues given by
\beq\label{eq5}
\begin{aligned}
&\epsilon^\xi_{\lambda,\qq} = c^\xi_{\qq,0} + \lambda f^\xi_\qq, \\
&\vec u^{\,\xi}_{\lambda,\qq} = \frac{g^\xi_\qq}{f^\xi_\qq\sqrt{2w^\xi_{\qq,\lambda}}}\left(\begin{array}{c}\frac{f^\xi_\qq\,w^\xi_{\qq,\lambda}}{g^\xi_\qq}e^{-i/2 \varphi^\xi_\qq} \\e^{i/2 \varphi^\xi_\qq}\end{array}\right),
\end{aligned}
\eeq
where $f^\xi_\qq=\sqrt{\left(g^\xi_\qq\right)^2+\left(c^\xi_{\qq,3}\right)^2}$, $w^\xi_{\qq,\lambda}=1+\lambda\left(c^\xi_{\qq,3}/f^\xi_\qq\right)$ (with $\lambda=\pm 1$ being the band index, identifying the valence ($\lambda=-1$) or conduction ($\lambda=+1$) band of the material).

Notice, that this form of eigenvalue problem is fairly general, and it applies to every material (or physical process), whose low energy Hamiltonian can be parametrised with a $2 \times 2$ matrix. 
\section{Temporal and Spatial Inversion Symmetries of 2D Weyl Materials}
In this section we discuss the behaviour of the Hamiltonian defined above under time reversal and spatial inversion symmetry. In particular, we show how the former is preserved, while the latter is broken, due to the presence of the term $c_{\vett{q},3}^{\xi}$.
\subsection{Time Reversal Symmetry}
Let us start with time reversal symmetry. To prove that the Hamiltonian in Eq. \eqref{Hty2.1} preserves time reversal symmetry, we need to prove that, under the action of the time reversal operator $\hat{T}=\mathcal{K}$ (which, essentially, amounts to complex conjugation), the Hamiltonian in Eq. \eqref{Hty2.1} transforms as $\hat{T}H_{\xi}(\vett{q})=H_{-\xi}(\vett{q})$. An explicit calculation using Eqs.\eqref{Hty2.1}, and \eqref{cCoefficients} gives
\barr\label{eq6}
\hat{T}H_{\xi}(\vett{q})&\equiv&H^*_{\xi}(-\vett{q})\nonumber\\
&=&-\xi\sigma_0\left(\vett{a}\cdot\vett{q}\right)-\xi\sigma_x\left(\uvett{v}\cdot\vett{q}\right)_1\nonumber\\
&+&\sigma_y\left(\uvett{v}\cdot\vett{q}\right)_2+\sigma_z\left(-\xi v\,q_x+\frac{\Delta}{2}\right)\nonumber\\
&=&H_{-\xi}(\vett{q}),
\earr
where $(\uvett{v}\cdot\vett{q})_k$ indicates the $k$-th element of the vector $\uvett{v}\cdot\vett{q}$, and  we have made use of the relation $-\sigma_y^*=\sigma_y$. Notice, how to go from the third to the fourth line, we only made the substitution $\xi\rightarrow-\xi$, as it is expected from time reversal symmetry to hold. 
\subsection{Spatial Inversion Symmetry}
For the case of spatial inversion symmetry, the Hamiltonian in Eq. \eqref{Hty2.1} should transform, under the action of the inversion operator $\hat{I}=\sigma_x$, as follows
\beq 
\hat{I}\,H_{\xi}(-\vett{q})\,\hat{I}^{\dagger}=H_{-\xi}(\vett{q}).
\eeq 
Again, using Eqs. \eqref{Hty2.1} and \eqref{cCoefficients}, we can show that this symmetry is broken by calculating explicitly the left- and right-hand side of the above equality and show they give a different result. In partcular, we have that
\barr 
\hat{I}\,H_{\xi}(-\vett{q})\,\hat{I}^{\dagger}&=& -\xi\,\sigma_0\,\left(\vett{a}\cdot\vett{q}\right)\nonumber\\
&-&\xi\,\sigma_x\,\left(\uvett{v}\cdot\vett{q}\right)_1+\sigma_y\,\left(\uvett{v}\cdot\vett{q}\right)_2\nonumber\\
&+&\sigma_z\,\left(\xi\,v\,q_x-\frac{\Delta}{2}\right).
\earr 
A direct comparison of the result above with Eq. \eqref{eq6} allows us to immediately conclude, that $\hat{I}\,H_{\xi}(-\vett{q})\,\hat{I}^{\dagger}\neq\,H_{-\xi}(\vett{q})$, and therefore that spatial inversion symmetry is broken. Notice, how this symmetry is broken because of the presence of the last term, proportional to $\sigma_z$, which corresponds to the $c_{\vett{q},3}^{\xi}$ term in Eq. \eqref{Hty2.1}.
\section{Dirac-Bloch Equations for 2D Weyl materials}
We can now use the formalism developed so far to derive the Dirac-Bloch equations (DBEs) for a generalised 2D Weyl material. To do so we introduce the minimal substitution $\qq \to \boldsymbol{\pi}(t) = \qq + e \, \A(t)$, where $\A(t)$ is an external electromagnetic field.  The Hamiltonian is now explicitly time dependent, but it can be still diagonalised in term of instantaneous band eigenstates and eigenvalues. 

For the purposes of this work, we employ a linearly polarised, spatially homogeneous, Gaussian electromagnetic pulse, whose vector potential has the following explicit form
\beq 
\vett{A}(t)=E_0\,\tau\,e^{-t^2/\tau^2}\cos(\omega_L t)\,\uvett{f}
\eeq 
where $E_0$ is the electric field amplitude, $\tau$ the duration (FWHM) of the pulse, $\omega_L$ its carrier frequency, anf $\uvett{f}=\{\uvett{x},\uvett{y}\}$ indicates the pulse's polarisation, which we choose to be either oriented along the $x$-  or $y$-axis.

Next, we can represent the time dependent, momentum space coefficients $c_{\boldsymbol\pi,\mu}^{\xi}(t)$, in direct space by means of the formal substitution (i.e., inverse Fourier transform)$\vett{q}\,\rightarrow\, -i\hbar\,\nabla$, to obtain the following time-dependent \emph{Type-II Weyl equation}
\beq
i \hbar \Pdert \vec \psi^{\, \xi}(\rr,t) = \sum_{\mu} \sigma_\mu \mathcal D_\mu\, \vec \psi^{\, \xi}(\rr,t),
\eeq
where the differential operators $\mathcal D_\mu$ have the following explicit expression
\beq
\begin{aligned}
\mathcal D_0 &= a_x D_x + a_y D_y,\\
\mathcal D_1 &= v_{xx} D_x + v_{xy} D_y,\\
\mathcal D_2 &= v_{yx} D_x + v_{yy} D_y,\\
\mathcal{D}_3&=v D_x+\frac{\Delta}{2},
\end{aligned}
\eeq
where $D_\mu = \partial_\mu + eA_\mu(t)/\hbar c$ is the standard covariant derivative \cite{maggiore_modern_2005}. Note, that while $D_{\mu}$  transforms as a three-vector under the Lorentz group, $\mathcal{D}_{\mu}$ does not. To prove this, we can define a general form of $\mathcal{D}_{\mu}$ from the relations above, as linear combination of covariant derivatives along $x$ and $y$ with some constant coefficients, i.e., $\mathcal{D}_{\mu}=a_{\mu}\, D_x+b_{\mu}\,D_y$, where $a_{\mu},b_{\mu}$ can be determined by using the relations above. Applying a Lorentz transformation $\Lambda^{\nu}_{\mu}$ to $\mathcal{D}_{\mu}$ then gives the following result
\barr 
\mathcal{\tilde{D}}_{\mu}&=&a_{\mu}\,\tilde{D}_x+b_{\mu}\,\tilde{D}_y\nonumber\\
&=&a_{\mu}\,\left(\Lambda^{-1}\right)^{\nu}_x D_{\nu}+ b_{\mu}\,\left(\Lambda^{-1}\right)^{\nu}_y\,D_{\nu}\nonumber\\
&=&\left[a_{\mu}\,\left(\Lambda^{-1}\right)^x_x+b_{\mu}\,\left(\Lambda^{-1}\right)^x_y\right]\,D_x\nonumber\\
&+&\left[a_{\mu}\left(\Lambda^{-1}\right)^y_x+b_{\mu}\,\left(\Lambda^{-1}\right)^y_y\right]D_y\nonumber\\
&\neq&\left(\Lambda^{-1}\right)^{\nu}_{\mu}\mathcal{D}_{\nu},
\earr 
Hence, the equation above is not Lorentz covariant. 

The general spinor solution can be then expressed in term of instantaneous band eigenstates as 
\beq \label{spinor}
 \vec \psi_\xi(\rr,t) =\frac{1}L \sum_{\qq,\lambda} \, \hat a_{\qq,\lambda} \vec u^{\xi}_{\lambda,\qq} e^{-i(\tilde \omega_{\qq,\lambda}(t)+\gamma_{\qq,\lambda}(t)-\qq \cdot \rr)},
 \eeq
where $\hat a_{\qq,\lambda}$ are the band-electrons ladder operators for valence ($\lambda=-1$) and conduction ($\lambda=+1$) bands, respectively, 
\beq 
\tilde \omega_{\qq} (t) = \frac{\lambda}\hbar \int_0^t\,d\tau\, \epsilon_{\qq} (\tau)
\eeq 
is the dynamical phase, and
\beq 
\gamma_{\qq,\lambda}(t)=\int_{-\infty}^t\,d\tau\,\Omega_{\lambda\lambda}(\tau)
\eeq 
is the Berry phase, with $\Omega_{\lambda\mu}(t)$ being the intraband (if $\lambda=\mu$) or interband (if $\lambda\neq\mu$) Rabi frequency, defined as $\Omega_{\lambda\mu}(t)=\left(\vec u^{\,\xi}_{\mu,\qq}\right)^{\dagger}\cdot\vec u^{\,\xi}_{\lambda,\qq}$.

Inserting the Ansatz in Eq. \eqref{spinor} into the field Hamiltonian 
\beq
H = \int d\rr \, \vec \psi_\xi^{\,\dagger}(\rr,t) \sigma_\mu \mathcal D^\mu \vec \psi_\xi(\rr,t),
\eeq
and upon introducing the electron-hole ladder operators $\hat a_{\qq,1} = \hat a_\qq$ and $\hat a_{\qq,-1} = \hat b^\dagger_{-\qq}$, and from the Heisenberg equations of motion for these ladder operators, we arrive, after a lengthy but straightforward calculation, at the following set of DBEs for the population and inversion variables $n^e_{\qq} = \braket{\adq \haq}$, $n^h_{\kk}=\braket{\bdq \hbq}$ and $p_\qq = \braket{b_{-\qq} a_\qq}$, i.e.,
%
\beq \label{RDB}
\begin{cases}
\hbar \dot p_{\qq}^{\xi} &= -2i f_{\qq}^{\xi}(t) p_{\qq}^{\xi} - i \hbar \Omega_{\qq}^{\xi}(t)\,e^{-2i\gamma_{\qq}^{\xi}(t)}\Delta n_{\vett{q},\xi}, \\
\\
\dot n^{e,h}_{\qq} &= - 4\operatorname{Re}\left\{\left(\Omega_{\qq}^{\xi}(t)\right)^*\,e^{2i\gamma_{\qq}^{\xi}(t)}p_{\qq}^{\xi}\right\},
\end{cases}
\eeq
where $\Delta n_{\vett{q},\xi}= (n_{\qq,\xi}^{e} + n_{\qq,\xi}^{h} - 1)$, and $\Omega_{\qq}^{\xi}(t)\equiv\Omega_{\lambda,-\lambda}(t)$. An important aspect of the DBEs is that while describing the interband transitions they encapsulate the intraband dynamisc trough the instantaneous energy and dynamical angle $\varphi_\qq^\xi(t)$, as well as the Berry phase $\gamma_{\qq}(t)$. In the following section we compute explicitly the interband and intraband contributions to the nonlinear current.
\section{Generalised Nonlinear current}
We now compute explicitly the time dependent current that we will use to study the nonlinear response of Type-II 2D-WMs. The quantised current density can be written as
\beq
\boldsymbol{\mathcal{J}}= -e\sum_\xi \vec\psi^{\, \dagger}_\xi \,\nabla_\pi H^\xi(\qq) \,\vec\psi_\xi,
\eeq
whose components read
\beq 
J_{\mu}=-e\,\sum_{\xi,\mu}\,\psi^{\dagger}_\xi\,\sigma_{\mu}\partial_{\pi_{\mu}}c_{\vett{q},\mu}^{\xi}\vec\psi_\xi.
\eeq
To obtain a simple explicit expression for the current copmonents $J_{x,y}$, it is useful to compute first quantities of the form $\vec\psi^{\dagger}_\xi\sigma_\mu\vec\psi$. To do so, we use Eqs. \eqref{eq5} to write the products in terms of the instantaneous eigenstates $\vec u_{\lambda,\vett{q}}^{\xi}(t)$ and then compute the following quantities
\beq
\begin{aligned}
&\vec u^{\,\xi \dagger}_{\lambda,\qq}(t) \sigma_0 \vec u^{\,\xi}_{\lambda',\qq}(t)  = \delta_{\lambda,\lambda'},\\
&\vec u^{\,\xi \dagger}_{\lambda',\qq}(t) \sigma_x \vec u^{\,\xi}_{\lambda,\qq}(t) = \begin{cases} z^\xi_\qq \cos\varphi^\xi_{\qq}(t) - i \lambda \sin \varphi^\xi_{\qq}(t) \quad &\lambda' = -\lambda \\  \lambda \frac{g^\xi_\qq}{f^\xi_\qq}\cos \varphi^\xi_{\qq}(t) \quad &\lambda' = \lambda ,\end{cases}  \\
&\vec u^{\,\xi \dagger}_{\lambda,\qq}(t) \sigma_y \vec u^{\,\xi}_{\lambda',\qq}(t) = \begin{cases} z^\xi_\qq \sin \varphi^\xi_{\qq}(t) + i\lambda \cos \varphi^\xi_{\qq}(t) \quad &\lambda' = -\lambda \\  \lambda \frac{g^\xi_\qq}{f^\xi_\qq}\sin \varphi^\xi_{\qq}(t) \quad &\lambda = \lambda',
\end{cases} \\
&\vec u^{\,\xi \dagger}_{\lambda,\qq}(t) \sigma_z \vec u^{\,\xi}_{\lambda',\qq}(t)  = \begin{cases} -\lambda z^\xi_\qq \quad &\lambda' = -\lambda , \\ \frac{g^\xi_\qq}{f^\xi_\qq} \quad &\lambda' = \lambda,
\end{cases}
\end{aligned}
\eeq
where, according to the generalised polar coordinates introduced in Eqs. \eqref{polarCoord}
$\cos \varphi^\xi_\qq(t) =c^\xi_{1,\qq}(t)/g_\qq^\xi(t)$ and $\sin\varphi^\xi_\qq(t)=c^\xi_{2,\qq(t)}/g_\qq^\xi(t)$. To make the notation more compact we have also defined the function $z^\xi_\qq = -c^\xi_{\qq,3}/f^\xi_\qq$. Using the above results, the components of the nonlinear current operator can be then written in the following simple form, as a function of the coefficients $c_{\vett{q},\mu}^{\xi}$
\barr\label{nonlinearCurrent}
J_\mu(t) &= &-e\sum_{\xi,\qq}\Biggl( \partial_{\pi_\mu} c^\xi_{\qq,0} +  \partial_{\pi_\mu} c^\xi_{\qq,1} \frac{c^\xi_{\qq,1}}{f^\xi_{\qq}} +  \partial_{\pi_\mu} c^\xi_{\qq,2} \frac{c^\xi_{\qq,2}}{f^\xi_{\qq}}\nonumber\\
&-&   \partial_{\pi_\mu} c^\xi_{\qq,3} z^\xi_\qq\Biggl) (n^e_\qq + n^h_\qq - 1) + 2\text{Re}\Bigg\{\Biggl[\partial_{\pi_\mu} c^\xi_{\qq,1}\nonumber\\
&\times&\Biggl(z^\xi_\qq \frac{c^\xi_{\qq,1}}{g^\xi_{\qq}} - i \frac{c^\xi_{\qq,2}}{f^\xi_{\qq}}\Biggl) + \partial_{\pi_\mu} c^\xi_{\qq,2} \Biggl(z^\xi_\qq \frac{c^\xi_{\qq,2}}{g^\xi_{\qq}} + i \frac{c^\xi_{\qq,1}}{f^\xi_{\qq}}\Biggl)\nonumber\\
&+& \partial_{\pi_\mu} c^\xi_{\qq,3} \frac{g^\xi_\qq}{f^\xi_\qq} \Biggl] p_\qq e^{-2i\gamma^\xi_{\qq}(t)}\Bigg\} 
\earr
\section{Nonlinear Optical response of 2D Weyl Materials}
The nonlinear optical response of material presenting either a Dirac or Weyl low energy dispersion has been the focus of relevant research in recent years. In particular, many studies focused on the role of the nodal point, both gapped or gapless, in the interaction with an external electromagnetic field  \cite{Ishikawa_2010,cb,cm,lw,gd,yj,dj,ike}. Remarkably it has been pointed out that the spike-like Berry curvature, in type-I and type-II bulk Weyl materials, plays an important role in the generation of efficient high order harmonic generation \cite{lw,gd,yj,dj}.
2D-WMs have not been observed in nature yet, nor have they been fabricated. Specifying a set of parameters $\{\vett{a},\hat{\vett{v}}\}$ that would correspond to a specific 2D-WM, therefore, is not possible. Tilted DW cones, however, have been observed in several bulk and layered materials, such as Mo$\text{Te}_2$, TaIr$\text{Te}_4$, and W$\text{Te}_2$ \cite{LSA2021},  the latter also showing signatures of tilted Weyl cones in 2D monolayers \cite{natPhys2018}, or layered organic conductors, which possesses Dirac, rather than Weyl, tilted cones \cite{hirata_observation_2016}. Following the discussion presented in these references, however, allows us to make reasonable assumptions on the shape and values of the material parameters. 

In our simulations, therefore, for the Weyl material we assume a diagonal form of the velocity tensor with $v_{xx}=6.70\,v_0$, $v_{yy}=6.86\,v_0$, and a tilting vector $\vett{a}=(5.06,0.75)v_0$, where $v_0=0.01\,v_F$ and $v_F=c/300$ is the Fermi velocity in graphene \cite{katsnelson}. 

For the impinging electromagnetic field, on the other hand, we assume an amplitude of $E_0=10^7$ $V/m$, a carrier frequency of $\omega_L=12$ THz, and a pulse duration of $\tau=50$ fs.

To characterise the nonlinear optical response we solve Eqs. \eqref{RDB} in the vicinity of the nodal points using a Runge-Kutta-based solver, coded using the Julia package \emph{DifferentialEquations.jl} \cite{rackauckas2017differentialequations}, and then we compute the nonlinear electrical current from Eq. \eqref{nonlinearCurrent}. To do this, we exploit the axial symmetry of the system to perform the integration in momentum space in polar coordinates, which allows to achieve better accuracy in particular in the vicinity of the nodal points. We then calculate the nonlinear signal with the relation $I(\omega) \thicksim | \omega\, \mathbf{J}(\omega) |^2.$
\begin{figure}[!t]
    \centering
    \includegraphics[width=0.45\textwidth]{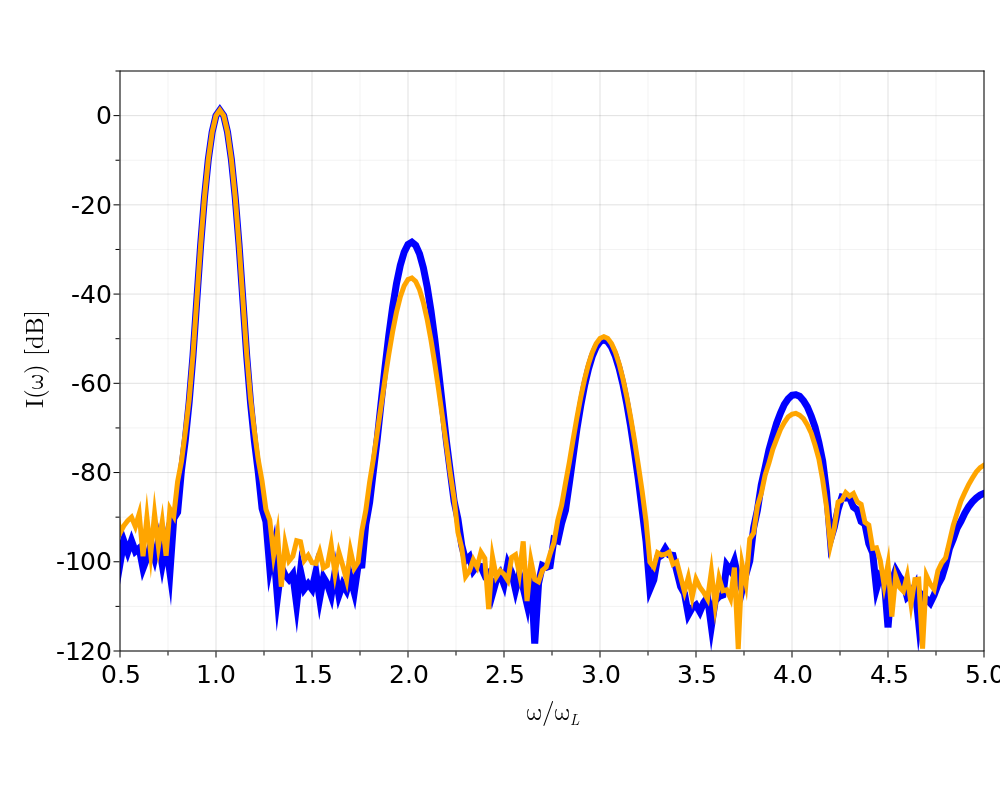}
    \caption{Nonlinear Optical Response $I(\omega)$ of a 2D-WM, as a function of normalised frequency $\omega/\omega_L$ for the case of an impinging electromagnetic pulse with $x$-polarisation (blue, solid line), and $y$-polarisation (orange, solid line). As it can be seen, the presence of tilted Weyl cones leads to the generation of even harmonics, and a polarisation-dependent amplification of the harmonic signal. The numerical values of the various parameters for these plots are given in the text.}
    \label{fig:nop_l_p}
\end{figure}

The nonlinear optical response of a 2D-WM for the case of impinging $x$- (blue line) and $y$-polarisation (orange line) is shown in Fig. \ref{fig:nop_l_p}. First, we notice that the generation of even harmonics (EHG) is not suppressed, as in the case of other 2D materials, such as graphene \cite{Ishikawa_2010}. This is mainly due to the presence of tilted, rather than straight, DW  cones. A nonzero tilting vector $\mathbf{a}$, in fact, breaks centrosymmetry, thus allowing the generation of even harmonics. 

The anisotropy induced by the tilting, and quantified by the tilting vector $\mathbf{a}$, also results in a polrisation-dependent amplification of the nonlinear signal, as it can be seen by comparing the spectra for $x$- and $y$-polarisation in Fig. \ref{fig:nop_l_p}. In particular, comparing the second-harmonic signal (SHS) for both polarisations, we can see that $I_{SHS}^{(x)}/I_{SHS}^{(y)}\simeq 10$ dB. We moreover notice, that while the polarisation dependence of the SHS (and, in general, the nonlinear response) of 2D-WMs is essentially due to the anisotropy induced by the tilting vector $\vett{a}$, the specific polarisation direction experiencing a higher SHS (or, more in general, a higher harmonic signal) is dictated by the actual form of the tilted cone, which, for the parameters chosen here, has a stronger tilt along the $x$-direction, than along the $y$-direction, resulting in an overall amplification of the nonlinear signal generated by $x$-polarised light. 

This is the main result of our work. The presence of tilted cones in 2D-WMs leads to a polarisation-dependent harmonic spectrum, which allows, on one hand, the generation of even harmonics and, on the other hand, presents an unbalanced response for different impinging polarisation states, which ultimately, is due to the orientation of the tilting vector $\vett{a}$. The great sensitivity of the nonlinear response of 2D-WMs to polarisation could be employed, for example, to realise efficient nanoscale nonlinear polarisation sensors. Interestingly, the strong anisotropy shown by the harmonic response of type-II Weyl fermions has been studied, for bulk materials, in two very recent experimental papers \cite{yj,dj}. These studies focus on the polarisation dependence of the nonlinear harmonic response of two experimentally confirmed type-II Weyl semimetal due to breaking of inversion-symmetry, $\beta$-WP$_2$ \cite{yj} and T$_d$WTe$_2$ \cite{dj}. It is worth noticing that the strong polarisation dependence shown by type-II fermions is not just remarkable if compared with their non-tilted counterparts \cite{cb,cm}. Twisted bilayer graphene, for example, shows both a nodal crossing point and anisotropy (due to the twisting angle), but while having different selection rules that depends on the photo-current direction, it does not show a relevant polarisation dependence for the intensity of the response \cite{ike}.
\begin{figure}[!t]
    \centering
    \includegraphics[width=0.45\textwidth]{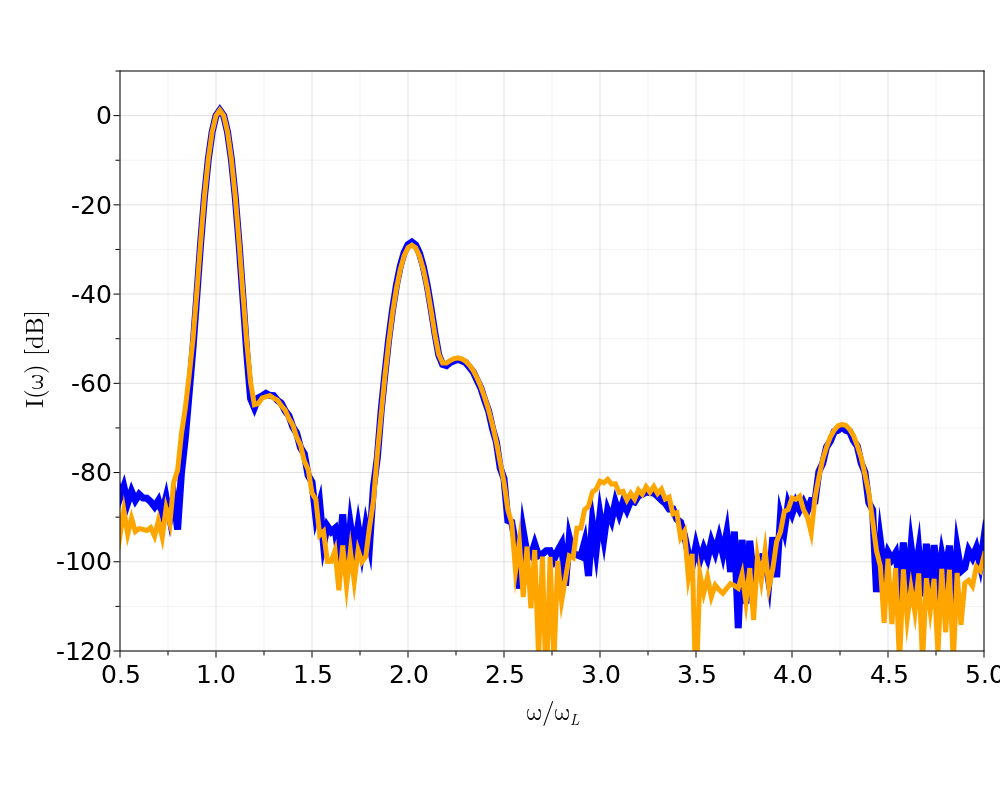}
    \caption{Frequency spectrum of the $x-$ (dashed, blue line) and $y-$component (solid, orange line) of the nonlinear current $\vett{J}(\omega)$ for the case of an impinging right-handed circularly polarised pulse.}
    \label{fig:nop_circular}
\end{figure}

Fig. \ref{fig:nop_l_p} also hints at another peculiar behaviour of the nonlinear response of 2D-WM. If we take a closer look at the odd harmonics, in fact, it is possible to notice that the amplification trend mentioned above for second-harmonic (and, in general, for even harmonics) is reversed, meaning that the odd harmonics generated by an impinging electromagnetic field polarised along the tilting direction of the DW cones are less amplified than those generated by a field with orthogonal polarisation (with respect to the tilting direction). This intrinsic feature of 2D-WM, essentially due to the strong anisotropy (regulated by the tilting vector $\textbf{a}$), would then allow to use polarisation as an active diagnostic tool to fully characterise the orientation of the DW cones in 2D-WM.

This unbalance, however, only manifests if the impinging electromagnetic pulse has linear polarisation. The tilted nature of DW cones, in fact, does not introduce any unbalance for circularly polarised pulses, as can be seen in Fig. \ref{fig:nop_circular}, where the spectral components of the nonlinear current generated by a right-handed circularly polarised pulses are plotted as a function of the harmonics of the impinging pulse. As it can be seen, the spectra for the $x$ (dashed, blue line in Fig. \ref{fig:nop_circular}), and $y$ (solid, orange line in Fig. \ref{fig:nop_circular}) are very similar, and do not present the characteristic unbalance seen in Fig. \ref{fig:nop_l_p}. The effect of the tilting, in this case, can be seen by the appearance of even harmonics, not present for the case of a straight cone, but no difference in the spectral amplitude of the two components of the nonlinear current can be seen.

%
%
%
%
%
\begin{figure}[!t]
    \centering
    \includegraphics[width=0.45\textwidth]{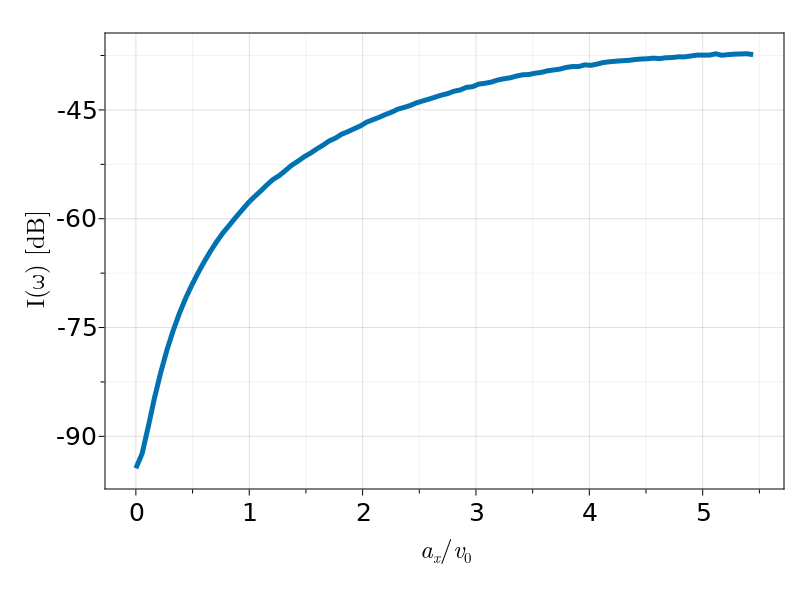}
    \caption{Evolution of the peak of the SHS, for the case of Fig.\ref{fig:nop_l_p}, as a function of the scaled tilting parameter $a_x/v_0$.}
    \label{fig:nop_a_y}
\end{figure}

\emph{Tuning the nonlinear response of 2D-WMs - }An important aspect to look at, when considering the potential impact 2D-WMs could have for the realisation of different linear and nonlinear photonic devices, is to understand how their different degrees of freedom, such as anisotropy and tilting, can be engineered to optimise the onset of desiderable features, like the polarisation-dependent SHS amplification described above. To this aim, in Fig. \ref{fig:nop_a_y} we show how changing the tilting vector $\vett{a}$ along the $x$-direction (while maintaining it constant in the $y$-direction) affects the intensity of the SHS. As it can be seen, varying the tilting parameter $a_x$ corresponds to a nonlinear modulation of the intensity of the SHS, which saturates to approx. 60 dB for tilting values $a_x\geq 5$. This, moreover, is not limited to SHS, and the same kind of analysis could be applied, in principle, to optimise the material to amplify a specific harmonic signal. Suitably engineering the tilting vector $\mathbf{a}$ of 2D-WMs could then lead to very efficient, polarisation-sensitive frequency converters. Nonlinear frequency conversion is a useful tool for the manipulation of laser light. Optical non-linearity is utilised for converting part of the optical power of the input light to output light in a different wavelength region \cite{LLi,bk}. In this perspective the anisotropy of the sample introduces further possibilities. It would allow to select the optimal polarisation of the input light for the efficiency of the conversion.
\section{Conclusions}
In this work, we have introduced a low-energy model for Type-II 2D-WMs, and studied its nonlinear optical response. In particular, we have show how the tilted nature of Type-II DW cones results in a natural polarisation-dependent anisotropy of the nonlinear signal, and in a significant polarisation-dependent amplification (see Fig. \ref{fig:nop_l_p}). Moreover, we have also discussed how it is possible to control the nonlinear response of such materials, by suitably engineering its properties, in terms of velocity tensor and tilting vector. The results presented here suggest that 2D-WMs possess a significant potential for disruptive photonic devices, such as, for example, very efficient nonlinear polarisation sensors. In addition to that, our work could be impactful in the topological photonics community as well, where Type-II Dirac cones have recently made their appearance in photonic lattices \cite{natComm2020}. Understanding the formation and dynamics of Type-II tilted Weyl cones in such materials could, in fact, open new possibilities for the control of light in photonic nanostructures.
%

Y.T. and M.O. acknowledge the financial support of the Academy of Finland Flagship Programme, Photonics research and innovation (PREIN), decision 320165. Y.T. also acknowledges support from the Finnish Cultural Foundation, decision 00221008.
L D. M. V. acknowledges support from the European Commission under the EU Horizon 2020 MSCA-RISE-2019 programme (project 873028 HYDROTRONICS) and of the Leverhulme Trust under the grant RPG-2019-363.

\bibliography{DBbib}
\end{document}